# A Digital Twin Framework for Adaptive Treatment Planning in Radiotherapy

Chih-Wei Chang[1,*], Sri Sai Akkineni[2], Mingzhe Hu[1], Keyur Shah[1], Yuan Gao[1], Pretesh Patel[1], Ashesh B. Jani[1], Greeshma Agasthya[1,3], Jun Zhou[1] and Xiaofeng Yang[1,*]

[1]Department of Radiation Oncology and Winship Cancer Institute, Emory University, Atlanta, GA 30308

[2]Department of Medicine, Medical College of Georgia, Augusta, GA, 30912

[3]George W. Woodruff School of Mechanical Engineering, Georgia Institute of Technology, Atlanta, GA 30332

*Corresponding to: chih-wei.chang@emory.edu (CC) and xiaofeng.yang@emory.edu (XY)

**Running title:** Adaptive radiotherapy using digital twins

**Manuscript Type:** Original Research

**Abstract**

**Objective:** This work aims to develop a digital twin (DT) framework for fast online adaptive proton therapy planning in prostate stereotactic body radiation therapy (SBRT) with dominant intraprostatic lesion (DIL) boost, achieving clinical-equivalent plan quality with significantly reduced reoptimization time compared to traditional clinical workflows.

**Approach:** The proposed DT framework integrates deep learning–based multi-atlas deformable image registration, daily patient anatomy updates, and knowledge-based plan quality evaluation to enable predictive and adaptive radiotherapy. Leveraging a database of 43 prior prostate SBRT cases, the framework forecasts potential interfractional anatomical variations for a new patient and pre-generates multiple probabilistic treatment plans. Upon acquiring daily cone-beam CT (CBCT) for the new patient, the framework facilitates rapid online plan reoptimization. Plan quality is assessed using the ProKnow scoring system, evaluating dose coverage to the DIL and clinical target volume (CTV), as well as sparing of organs at risk (OARs).

**Main Results:** The DT framework achieved an average reoptimization time of 5.5 ± 2.7 minutes, generating the optimal DT-based plans with a plan quality score of 157.2 ± 5.6, matching or exceeding clinical plans. In contrast, clinical plans required 19.8 ± 11.9 minutes to attain a comparable score of 153.8 ± 6.0. DT-based plans provided DIL V100 coverage of 99.5% ± 0.6% and CTV V100 of 99.8% ± 0.2%, with reduced OAR doses, including bladder V20.8Gy of 11.4 ± 4.2 $cm^3$, rectum V23Gy of 0.7 ± 0.4 $cm^3$, and urethra D10 of 90.9% ± 2.3%, which were comparable to the clinical quality standard.

**Significance:** The proposed DT framework facilitates rapid, clinically comparable adaptive proton therapy planning, reducing reoptimization time while preserving or enhancing clinical plan quality. By addressing interfractional anatomical variations efficiently, it enhances treatment precision, reduces OAR toxicity, and supports real-time personalized radiotherapy, offering a transformative approach for prostate SBRT with DIL boost.

# 1 Introduction

Initially conceptualized by Glaessgen *et al.* (Glaessgen and Stargel, 2012), digital twins (DT) merge intricate multiphysics, multiscale, and probabilistic modeling to guide critical decisions in a complex system integration. The DT paradigm has been increasingly adopted and extended within the field of radiation oncology, potentially enabling personalized treatment strategies, real-time treatment data integration for outcome prediction, predictive diagnostics, clinical trial design, and streamlined treatment workflows for advanced radiotherapy (Mollica *et al.*, 2024; Sumini *et al.*, 2024; Shen *et al.*, 2024; Sadée *et al.*, 2025). Chaudhuri *et al.* (Chaudhuri et al., 2023) proposed a predictive DT framework to optimize patient-specific radiotherapy regimens for high-grade gliomas, addressing uncertainties in tumor behavior and treatment outcomes to improve therapeutic efficacy. Chang *et al.* (Chang *et al.*, 2025) demonstrated a DT framework to enhance treatment precision and efficiency by addressing geometrical uncertainties and optimizing dose conformity in prostate stereotactic body radiation therapy (SBRT). This methodology has found innovative applications in personalized healthcare, serving as a computational surrogate for patients (Björnsson *et al.*, 2019; Hormuth *et al.*, 2021). Such virtual models allow for the systematic evaluation of diverse treatment protocols to pinpoint the most effective course for an individual. Abdollahi *et al.* (Abdollahi *et al.*, 2024) introduced the concept of theranostic DT to personalize radiopharmaceutical therapies for prostate cancer treatment, avoiding risks of over- or under-dosing and suboptimal outcomes However, clinical decision-making in medicine is complex, involving symptoms, disease pathways, and pharmaceutical interactions (Wu *et al.*, 2022). These multifaceted elements often hinder the creation of a robust, data-driven DT model for healthcare. The application of DTs in medicine is still emerging, with Katsoulakis *et al.* (Katsoulakis *et al.*, 2024) proposing a comprehensive definition that emphasizes dynamic treatment simulations and proactive health outcome predictions. In this study, we refine the DT concept to focus on proton online adaptive radiotherapy (Albertini *et al.*, 2020; Paganetti *et al.*, 2021; Feng *et al.*, 2025), presenting a case study that showcases its transformative potential in enhancing the precision of radiation treatment design and implementation.

Prostate cancer is a major public health issue in the United States, representing roughly 30% of new cancer cases among men in 2023 and 2024 (Schaeffer *et al.*, 2023; Schaeffer *et al.*, 2024; Siegel *et al.*, 2024). Histopathological studies of prostate biopsy samples have shown that dominant intraprostatic lesions (DILs), the most aggressive cancerous foci within the prostate, are present in 83% of patients (Chen *et al.*, 2000). These lesions constitute the most biologically aggressive neoplastic foci within the prostate gland (Guimond *et al.*, 2019). These lesions are critical because they are linked to local recurrence and disease progression after radiotherapy, making them a key target for treatment (Guimond *et al.*, 2019; Aizawa *et al.*, 2024). However, DIL's small size and anatomical shifts between treatment fractions complicate precise radiation delivery in external beam radiotherapy (EBRT), risking underdosing the tumor or damaging nearby tissues like the rectum or bladder (Lee *et al.*, 2019). The clinical importance of DILs stems from their role in treatment failure. Studies show that boosting radiation doses to DILs using techniques like SBRT can enhance tumor control probability without significantly elevating toxicity, provided DILs are not near sensitive structures (Kim *et al.*, 2020; Murray *et al.*, 2020; Zhou *et al.*, 2022). Clinical trials and meta-analyses further support that DIL-focused dose escalation improves local control and may extend disease-free survival, with toxicity profiles similar to standard whole-gland radiotherapy when strict organ-at-risk constraints are applied (von Eyben *et al.*, 2016; Herrera *et al.*, 2019). However, challenges persist, as higher boost doses or DILs located near critical areas can increase genitourinary or gastrointestinal toxicity risks, necessitating meticulous planning (Skrobala *et al.*, 2023).

Treating DILs with EBRT is particularly difficult due to their small volume and the prostate's susceptibility to interfractional anatomical variations, such as organ motion or deformation caused by bladder or rectal filling (Zhao *et al.*, 2025a; Zhao *et al.*, 2025b). These changes can shift DIL positions between sessions, complicating accurate targeting. Moreover, the proximity of DILs to critical structures often limits the feasibility of dose escalation, as even minor misalignments can lead to severe side effects (Andrzejewski *et al.*, 2015). These complexities underscore the need for innovative approaches to improve treatment precision and efficacy. Proton online adaptive radiotherapy with DT can potentially offer a promising solution to address these challenges in prostate cancer radiotherapy. By integrating multiphysics simulations and real-time data assimilation, a DT framework can dynamically model DIL behavior and anatomical variations, enabling precise radiation dose optimization (Kapteyn *et al.*, 2021). Specifically, a radiotherapy-focused DT can simulate treatment strategies, predict DIL response, and adjust for interfractional changes, ensuring accurate dose delivery while minimizing toxicity. This approach facilitates personalized treatment planning, computationally testing multiple regimens to identify the optimal strategy for each patient, thus overcoming the limitations of traditional EBRT in targeting small, mobile DILs.

In this study, we propose a DT framework to enhance adaptive proton therapy by integrating advanced proton treatment planning (Zhou *et al.*, 2022; Chang *et al.*, 2023), deep learning (DL)–based multi-atlas deformable image registration (DIR) (Yang *et al.*, 2014; Krebs *et al.*, 2019; Dubost *et al.*, 2020), and knowledge-based plan quality evaluation methods (Chang *et al.*, 2024; Gao *et al.*, 2024). The framework enables rapid online plan reoptimization using daily cone-beam computed tomography (CBCT) data. By predicting interfractional anatomical changes through DL-based DIR, the DT framework can pre-generate a set of probabilistic treatment plans. On each treatment day, the most suitable plan is selected based on the patient's current anatomy, streamlining the online adaptive workflow. Compared to conventional online reoptimization, this approach significantly reduces planning time and improves overall efficiency. This study demonstrates the application of a DL-enabled DT framework for proton online adaptive radiotherapy, with two key contributions aimed at evaluating its clinical feasibility:

- **Fast Online Plan Reoptimization**: The framework leverages DL-based DIR to integrate the DT concept into online proton adaptive therapy, enabling rapid plan reoptimization based on daily patient anatomy. By utilizing existing plan conditions, it eliminates the need for time-consuming multicriteria optimization.
- **Efficient Plan Selection and Tissue Sparing**: The framework provides pre-approved probabilistic treatment plans tailored to predicted anatomical changes, ensuring DIL dose coverage, minimizing healthy tissue exposure, and simplifying adaptative treatment without same-day quality assurance CT (QACT) scans.

This predictive approach streamlines the creation of online treatment plans by leveraging historical data, enabling rapid, precise, and personalized adaptive therapy.

## 2 Materials and methods

### 2.1 A digital twin framework for online adaptive proton treatment

The application of DT technology in proton radiotherapy is in its early stages, yet it holds transformative potential for precision medicine. Sadée *et al.* (2025) addressed the absence of a consensus on medical digital twins, defining their five essential components, patient, data connection, patient-in-silico, interface, and twin synchronization, and exploring enabling technologies and implementation strategies. The article highlights the ability of medical DT to enhance health outcomes and reduce clinical workloads. A key component, the patient-in-silico, is a high-fidelity virtual model that integrates mechanistic modeling and artificial intelligence to predict treatment outcomes and support clinical decision-making. Building on this framework, we propose a novel DT-based approach, illustrated in Figure 1, to advance adaptive proton therapy, including three modules: the multi-atlas DIR module, clinical proton treatment workflow, and knowledge-based plan quality evaluation module to ensure the optimal treatment plan delivery. The multi-atlas DIR module leverages DL-based DIR and daily CBCT to enable rapid, online adaptive proton therapy for prostate SBRT with DIL boost, optimizing treatment precision and efficiency. A conventional clinical workflow for online treatment evaluation using daily CBCT, proceeding with treatment if dosimetric parameters align with planning conditions. Conversely, a decline in DIL or CTV coverage during online dose evaluation suggests anatomical changes or setup uncertainties compromising plan quality, necessitating intervention. In such cases, significant anatomical shifts trigger the multi-atlas DIR module for online plan adaptation using predicted CT (pCT), outlined in the red dashed box in Figure 1.

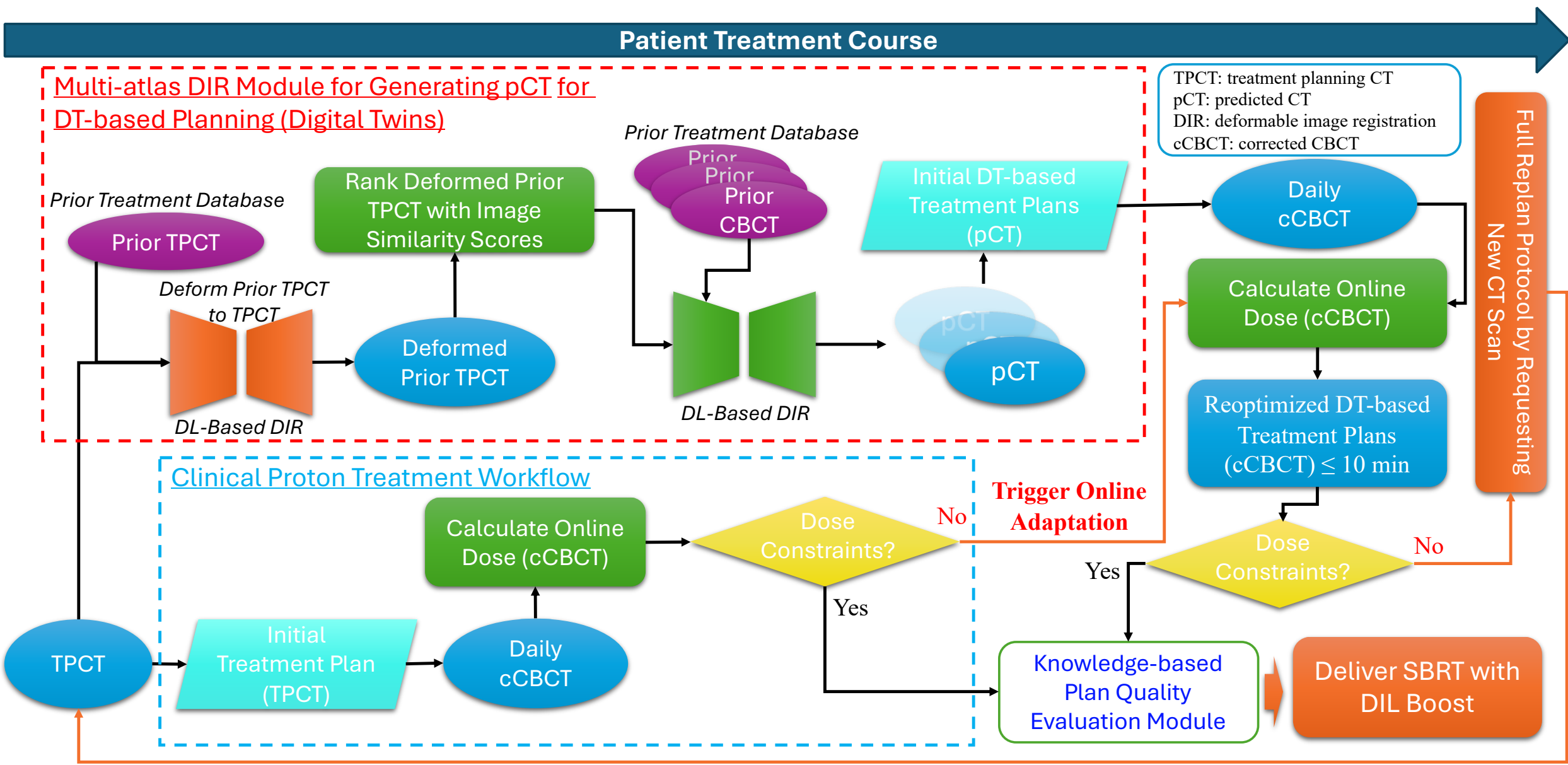

**Figure 1.** Digital twin (DT) framework for fast proton online treatment adaptation. The treatment planning incorporates predicted interfractional anatomical variations through predicted CT (pCT) image sets generated by deep learning (DL)-based deformable image registration (DIR), enabling rapid and personalized plan adaptation.

The framework's clinical workflow module supports initial planning using treatment planning CT (TPCT) and online dose evaluation with corrected daily (cCBCT) to account for dosimetric effects caused by interfractional anatomical changes (Chang *et al.*, 2023). The cCBCT enhances CBCT image quality to achieve Hounsfield units consistent with TPCT, enabling accurate dose calculations. If the dosimetric analysis aligns well with the initial treatment plan, the plan quality score is further evaluated using the knowledge-based plan quality evaluation module (Figure 1). This module enables clinicians to assess plan quality through multiple correlation functions, grading it based on dosimetric factors from the DIL, clinical target volume (CTV), and OARs. The module generates a composite score to inform clinicians about the current treatment quality, allowing comparison with prior treatments and updating the grading functions based on newly observed treatment data. Section 2.6 provides a detailed description of the correlation functions within the plan quality evaluation module. If the online dose evaluation indicates unsatisfactory dose constraints, online adaptation is triggered, incorporating DT-based treatment plans using pCT generated from the multi-atlas DIR module in Figure 1.

The multi-atlas DIR module, functioning as a patient-in-silico model, begins by inputting the TPCT. A DL-based DIR model, VoxelMorph (Balakrishnan *et al.*, 2019), aligns prior TPCTs from the institutional database with the current input TPCT. Image similarity between these prior TPCTs and the current TPCT is quantified using evaluation metrics described in Section 2.5. The top 20% of prior TPCTs, based on composite similarity scores, are selected to analyze motion correlations with their corresponding prior CBCTs, leveraging a data-driven approach to infer probable motion patterns from historical daily CBCT data. We hypothesize that biomechanical motion follows physiological constraints, ruling out arbitrary shifts. The deformation vector fields (DVFs) derived from these patterns are applied to the current TPCT to predict prostate motion, generating multiple pCT image sets for treatment planning. As this is a virtual process using a patient-in-silico approach, the resulting treatment plans, termed DT-based plans, incorporate potential interfractional anatomical changes. These DT-based pCT plans will serve as prior conditions for online reoptimization with daily CBCT. On the treatment day, if suboptimal dosimetric parameters necessitate online adaptation, the treatment plans using pCT, generated from the multi-atlas DIR module in Figure 1, enable rapid reoptimization based on daily CBCT. By selecting available, relevant, and adequately evaluated (AREAD) pCTs, the module ensures consistent prior planning conditions, facilitating clinically viable plan delivery within 10 minutes. We limit online reoptimization time for treatment planning to under 10 minutes to optimize clinical resources and prioritize patient comfort, reducing motion uncertainty while patients remain in treatment positions. In rare cases where neither DT-based nor clinical plans derived from the original TPCT are adequate, a full replanning protocol is initiated, requiring a new TPCT acquisition.

### 2.2 Patient data

The institutional CBCT image data were used in this work. Since our institution has not yet adopted prostate proton SBRT with two treatment fractions, we performed retrospective investigation using CBCT from 43 patients treated with five-fraction prostate SBRT. Of these, five patients received DIL boost treatment. Each patient underwent five cone-beam scans using the Varian CBCT imaging system from the proton ProBeam® machine, resulting in 210 CBCT image sets. The CBCT images were acquired at 125 kVp and 176 mA, reconstructed with a Ram-Lak kernel at a resolution of 1.0×1.0×2.0 mm³ with axial 104 slices. The DT framework was tested on the five patients, treated with SBRT and DIL boost. Leave-one-patient-out training was used to train DL-based DIR models for each patient with DIL for prediction of interfractional anatomical changes. A Siemens SOMATOM Definition Edge scanner was used to acquire patients' images from for initial treatment planning.

### 2.3 Two-fraction prostate SBRT clinical parameters

To the authors' knowledge, no clinical trials specific to proton-based two-fraction SBRT have published results. Consequently, this study adopts dose constraints from photon-based prostate SBRT trials (Ong *et al.*, 2023b). Table 1 shows the clinical parameters adapted from the 2SMART clinical trial (NCT03588819) (Ong *et al.*, 2023a) to implement the proposed DT framework for adaptive proton therapy planning. In the referenced trials, treatments were delivered with a frequency of once per week. The plan prescribed the CTV dose of 26 Gy with DIL boost of 32 Gy. The OAR dose constraints, including those for the bladder and rectum, are detailed in Table 1. In this study, prostate SBRT treatment plans were developed with DIL boost based on the 2SMART trial's prescription and clinical parameters outlined in Table 1. Radiation oncologists meticulously delineated and verified the contours of the CTV and OARs on both CT and CBCT images to ensure high accuracy and precision in dosimetric evaluations.

**Table 1.** Clinical parameters used in this retrospective DT-framework investigation for online adaptation of prostate SBRT with DIL boost.

| Parameters | Criteria |
| --- | --- |
| Prescription (DIL/CTV) | 32/26 Gy |
| CTV | V100 ≥ 98% |
| | D0.03cc < 110% |
| Urethra | D0.03cc < 105.6% |
| | D10% < 95% |
| Bladder | V14.6Gy < 25 cc |
| | V20.8Gy < 10 cc |
| Rectum | V13Gy < 7 cc |
| | V17.6Gy < 4 cc |
| | V23Gy < 1 cc |
| Online treatment planning | Reoptimization time < 10 min |

## 2.4 SBRT treatment planning

We used RayStation 2023B (RaySearch Laboratories in Stockholm, Sweden) to create efficient and precise proton therapy planning. The treatment planning system can generate the treatment quality CBCT by correct scatter and cavity artifacts and Hounsfield units to enable accurate radiation dose calculation (Thing *et al.*, 2022). The treatment planning system features GPU-based DIR and Monte Carlo (MC) dose calculations. This cCBCT enables rapid online treatment evaluation within approximately 2 minutes with extensive verification and validation based on proton water equivalent thickness and dose distributions using QACT (Chang *et al.*, 2023). Four proton beams with various gantry angles were used to design the prostate SBRT: bilateral (90° and 270°), left anterior oblique (50°), and right anterior oblique (310°) with the beam weighting of 35%, 35%, 15%, and 15%. Incorporating anterior-oblique beams minimizes radiation exposure to the rectum, as highlighted by Moteabbed *et al.* (Moteabbed *et al.*, 2017). The clinical prostate robust optimization includes 5 mm margins except for the posterior direction using 3 mm. The DT-based plans used a robust margin of 1.5 mm for positional uncertainty. A ±3.5% range uncertainty was used for all plans (Paganetti, 2012; Chang *et al.*, 2020), resulting in 21 optimization scenarios per plan to ensure treatment robustness.

## 2.5 Deep learning-based deformable image registration and image evaluation metrics

VoxelMorph (Balakrishnan *et al.*, 2019) is a DL-based framework aiming for fast medical DIR, designed to align pairs of images, such as 3D MRI brain scans or pelvic CBCT, by mapping them to a DVF using a convolutional neural network. Unlike traditional methods that optimize an objective function for each image pair, VoxelMorph parameterizes registration as a global function, optimized during training, enabling rapid registration. Its unsupervised learning approach, which relies solely on image intensities without requiring ground truth deformation fields, achieves accuracy comparable to state-of-the-art methods, as validated on a diverse dataset including pelvic CT and CBCT for prostate cancer radiotherapy (Hemon *et al.*, 2023; Wang *et al.*, 2023; Kolenbrander *et al.*, 2024). This unsupervised strength eliminates the need for cumbersome ground truth data, enhances generalizability, and supports scalability, while the framework's flexibility allows incorporation of auxiliary anatomical segmentations during training to further improve registration accuracy, making VoxelMorph a versatile tool for medical image analysis across anatomical sites.

Image evaluation metrics are critical for assessing the performance of medical image registration algorithms, each offering unique strengths. The Dice score (Dice, 1945) excels in measuring overlap between predicted and ground truth contours, ideal for evaluating anatomical structure alignment, as seen in VoxelMorph's pelvic registrations. HD95 score quantifies the 95th percentile Hausdorff distance (Huttenlocher *et al.*, 1993), emphasizing boundary accuracy and robustness to outliers in organ contours. Surface distance (SD) (Heimann *et al.*, 2009) evaluates average boundary discrepancies, providing precise surface alignment assessment. Peak signal-to-noise ratio (PSNR) (Korhonen and You, 2012) measures image reconstruction quality, highlighting pixel intensity fidelity in registered images. Intersection over Union (IoU) (Rezatofighi *et al.*, 2019) complements Dice by assessing segmentation overlap relative to the union, useful for imbalanced classes. Normalized surface distance (NSD) (Nikolov *et al.*, 2021) balances surface accuracy with tolerance, suitable for noisy segmentations. Accuracy (Naselaris *et al.*, 2015) gauges overall correctness of voxel classifications, effective for binary tasks. Precision (Maintz and Viergever, 1998) ensures low false positives, critical for avoiding over-segmentation, while Recall prioritizes low false negatives, ensuring comprehensive structure capture. Normalized cross-correlation (NCC) (Ayubi *et al.*, 2024) robustly measures intensity similarity across images with varying contrasts, as used in VoxelMorph's loss functions. Structural similarity index (SSIM) (Zhou *et al.*, 2004) captures perceptual image quality, emphasizing structural and textural alignment. Center of mass differences evaluate gross alignment by measuring centroid displacement between the reference and registered contours, useful for detecting large misregistrations. The learned perceptual image patch similarity (LPIPS) metric (Zhang *et al.*, 2018) leverages deep neural networks to assess perceptual similarity between images by comparing feature activations, offering a robust measure of visual quality that aligns closely with human perception. In this work, we sum all these metrics with equal weights to obtain the composite scores for a comprehensive ranking of image similarity in the proposed DT framework's multi-atlas DIR module (Figure 1), addressing overlap, boundary precision, intensity, and structural fidelity in medical imaging tasks.

### 2.6 Knowledge-based plan quality evaluation module using ProKnow scoring system

The suggested DT framework creates several potential treatment plans to ensure the best plan is delivered on the treatment day. Plan evaluations use cCBCT images taken on the treatment day to verify the precision of the patient's anatomical structure. To thoroughly evaluate plan quality, we employed the ProKnow® system (ProKnow Systems, Sanford, FL, USA) (Nelms *et al.*, 2012), a scoring approach previously applied in the 2016 AAMD/RSS-SBRT Prostate study (Richard Sweat *et al.*, 2016). We adapted the ProKnow system assigns scores to rank the treatment plans with different dose statistics of DIL, CTV, and OARs. The plans with higher scores indicate better quality while the plans with lower scores show least favorite dosimetry outcomes. Figure 2 displays our scoring system, which includes 9 scoring functions adapted from the original ProKnow functions to align with the data from two-fraction prostate SBRT clinical trials in Table 1. The dose metrics in Figure 2 include V100 (the volume percentage of the DIL/CTV received 100% prescription dose), bladder metrics of V14.6Gy and V20.8Gy, and rectum metrics of V13Gy, V17.6Gy, and V23Gy, urethra metrics of D0.03cc and D10 (doses to the most irradiated 0.03 $cm^3$ and 10% of the contour volume). Total plan quality is determined by combining the scores from each metric.

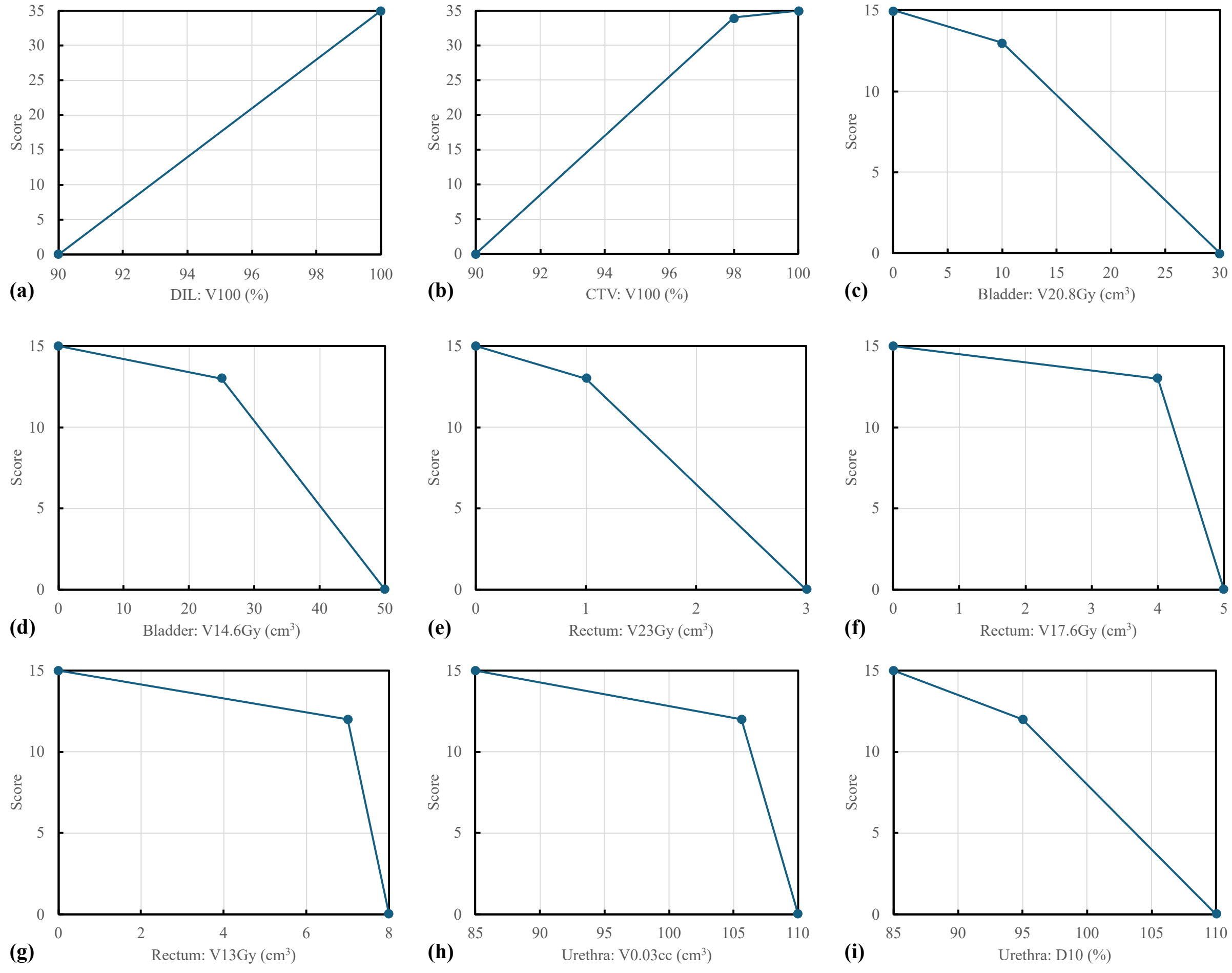

**Figure 2.** Scoring functions for assessing prostate SBRT plan quality. The scoring system evaluates cumulative doses from plans using daily CBCT, reflecting the patient's interfractional anatomy changes. (a)-(i) illustrate the scoring functions for various dose statistics.

## 3 Results

### 3.1 Treatment planning image set selection for two-fraction prostate SBRT with DIL

The proposed DT framework (Figure 1) generates pCT image sets for initial treatment planning, including a conventional clinical treatment plan based on TPCT and multiple DT-based treatment plans using motion-predicted pCT image sets for each patient. Each patient has five daily CBCT image sets. To evaluate the framework's performance, we selected two daily CBCT image sets with the lowest composite image similarity scores compared to TPCT, indicating the greatest interfractional anatomical changes. The composite score was calculated by equally weighting and summing SSIM, 1-LPIPS, and NCC. Table 2 presents image similarity comparison results of daily CBCT image sets (CB1 and CB2), selected to demonstrate two-fraction prostate SBRT with DIL, TPCT, and predicted pCT image sets for each patient.

**Table 2.** Image similarity comparisons of TPCT (Clinic) and pCT (DT-H and DT-L) against the treatment fraction 1 daily cCBCT (CB1) and treatment fraction 2 daily cCBCT (CB2). The pCT(DT-H) and pCT(DT-L) present the pCT with the highest and lowest similarity (composite scores) to the daily cCBCT from each treatment fraction.

| | Planning Image Set | SSIM | | 1 - LPIPS | | NCC | | Composite Score | |
|---|---|---|---|---|---|---|---|---|---|
| | | CB1 | CB2 | CB1 | CB2 | CB1 | CB2 | CB1 | CB2 |
| P1 | TPCT(Clinic) | 0.839 | 0.883 | 0.755 | 0.761 | 0.891 | 0.915 | 0.829 | 0.853 |
| | pCT(DT-H) | 0.865 | 0.868 | 0.771 | 0.766 | 0.895 | 0.878 | 0.843 | 0.837 |
| | pCT(DT-L) | 0.695 | 0.706 | 0.597 | 0.587 | 0.545 | 0.474 | 0.613 | 0.589 |
| P2 | TPCT(Clinic) | 0.847 | 0.839 | 0.766 | 0.787 | 0.909 | 0.919 | 0.841 | 0.848 |
| | pCT(DT-H) | 0.859 | 0.889 | 0.781 | 0.815 | 0.900 | 0.920 | 0.847 | 0.875 |
| | pCT(DT-L) | 0.677 | 0.685 | 0.606 | 0.626 | 0.660 | 0.702 | 0.648 | 0.671 |
| P3 | TPCT(Clinic) | 0.774 | 0.788 | 0.733 | 0.737 | 0.903 | 0.929 | 0.803 | 0.818 |
| | pCT(DT-H) | 0.766 | 0.779 | 0.707 | 0.724 | 0.864 | 0.870 | 0.779 | 0.791 |
| | pCT(DT-L) | 0.682 | 0.660 | 0.628 | 0.608 | 0.731 | 0.705 | 0.681 | 0.658 |
| P4 | TPCT(Clinic) | 0.719 | 0.805 | 0.665 | 0.757 | 0.826 | 0.936 | 0.737 | 0.833 |
| | pCT(DT-H) | 0.865 | 0.844 | 0.771 | 0.782 | 0.959 | 0.945 | 0.865 | 0.857 |
| | pCT(DT-L) | 0.701 | 0.666 | 0.606 | 0.581 | 0.698 | 0.744 | 0.668 | 0.664 |
| P5 | TPCT(Clinic) | 0.825 | 0.825 | 0.705 | 0.706 | 0.857 | 0.876 | 0.796 | 0.802 |
| | pCT(DT-H) | 0.824 | 0.862 | 0.705 | 0.719 | 0.796 | 0.866 | 0.775 | 0.816 |
| | pCT(DT-L) | 0.686 | 0.667 | 0.586 | 0.568 | 0.508 | 0.463 | 0.593 | 0.566 |

For the retrospective two-fraction prostate SBRT investigation, we identified CB1 and CB2 and calculated composite image similarity scores for these CBCT image sets against pCT image sets generated by the DT framework (Figure 1) for each patient. The pCT image sets were derived from the top 20% most similar prior TPCT images in the database, incorporating interfractional motion prediction. This selection yielded 9 prior patient imaging sets, resulting in 45 pCT images (5 prior CBCT images per prior TPCT). Figure 3 illustrates the image similarity comparisons of (a) CB1 and (b) CB2 against all 45 pCT image sets. Given the DT framework's ability to generate multiple pCT image sets based on anticipated anatomical variations and the extensive prior treatment database, we selected the most and least similar pCT images to demonstrate DT-based treatment planning. These selected pCT image sets, denoted pCT(DT-H) and pCT(DT-L) for high and low similarity relative to daily CBCT (CB1 and CB2) from treatment fractions 1 and 2, respectively, were used to generate SBRT treatment plans. Table 2 summarizes the image similarity metrics for pCT(DT-H) and pCT(DT-L).

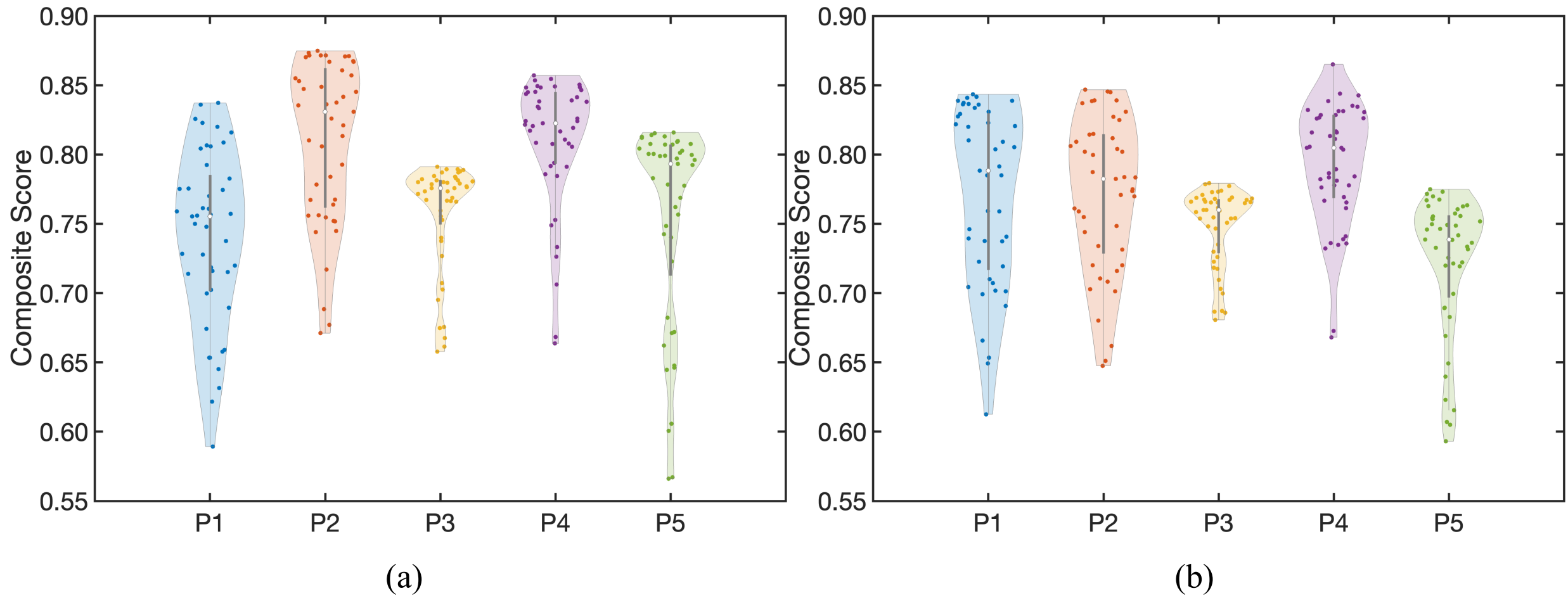

**Figure 3.** Image similarity comparisons of patients' (P1-P5) pCT image sets against (a) the treatment fraction 1 daily cCBCT (CB1) and (b) treatment fraction 2 daily cCBCT (CB2). Each patient includes 45 pCT image sets (5 prior CBCT images per prior TPCT for each patient).

### 3.2 Dose evaluation on daily CBCT

For treatment fraction 1 (Fx1), Table 3 presents a comparative summary of treatment plan (TP) evaluations for clinical, DT-H, and DT-L plans. The DT-H and DT-L plans were developed using pCT images generated by the DT framework, where "H" and "L" represent high and low similarity, respectively, between the pCT and the daily CBCT (CB1) from Fx1. The data in Table 3 reveal that clinical plans for all patients exhibit at least one dosimetric metric exceeding the dose constraints specified in Table 1, as assessed using daily CBCT for dose evaluation. This violation of dose constraints may initiate the online adaptation workflow. Notably, for Patients 2 and 4, the DIL V100 coverage falls below 90%, signaling a failure of DIL boost in delivering adequate DIL coverage during actual treatment. Table 4 provides dose evaluations based on the corrected CBCT image (CB2) from treatment fraction 2 (Fx2). Although clinical plans consistently achieve DIL V100 greater than 98% when evaluated with Fx2 daily CBCT, the OARs exceed the tolerances outlined in Table 1. Consequently, this breach necessitates the activation of the online adaptation process. Tables 3 and 4 also include online dose assessments using daily CBCT for DT-based plans prior to reoptimization under previous plan conditions. It is noteworthy that, for the DT-based plans (DT-H), Patients 3 and 5 meet the evaluation dosimetric criteria, DIL and CTV V100 > 95% and OARs < the constraints in Table 1, even before plan reoptimization.

**Table 3.** Dose volume endpoint comparisons of DIL, CTV, and OARs for prostate SBRT patients, obtained by online evaluation (Eval.) for the clinical plans (clinic) and the plans by the proposed DT framework based on the daily corrected CBCT acquired from the first treatment fraction images (CB1). Underlined numbers indicate dosimetric values that surpass the dose constraints outlined in Table 1, and can trigger online adaptation workflow.

| **CB1 Eval.** | TP | DIL (%) | CTV (%) | Bladder (cm$^3$) | | Rectum (cm$^3$) | | | Urethra (%) | |
|---|---|---|---|---|---|---|---|---|---|---|
| | | V100 | V100 | V20.8Gy | V14.6Gy | V23Gy | V17.6Gy | V13Gy | D0.03cc | D10 |
| P1 | Clinic | 98.7 | 99.9 | *15.4* | *29.7* | *1.7* | 3.6 | 6.2 | 92.3 | 88.9 |
| | DT-H | 98.9 | 99.7 | 13.4 | 25.9 | 1.5 | 4.1 | 7.3 | 90.0 | 87.4 |
| | DT-L | 98.8 | 99.3 | 13.2 | 25.9 | 1.9 | 4.6 | 8.3 | 91.4 | 88.6 |
| P2 | Clinic | *89.7* | *93.3* | *26.5* | *41.7* | *1.5* | 3.2 | 5.7 | 94.0 | 92.0 |
| | DT-H | 95.6 | 95.5 | 25.3 | 40.7 | 1.0 | 2.9 | 5.2 | 93.3 | 91.9 |

| | DT-L | 92.3 | 96.6 | 24.0 | 38.8 | 1.7 | 3.0 | 5.9 | 94.7 | 91.8 |
|---|---|---|---|---|---|---|---|---|---|---|
| P3 | Clinic | 100.0 | 99.7 | *12.0* | 21.6 | 0.2 | 2.1 | 4.4 | 103.1 | *99.1* |
| | DT-H | 99.6 | 99.7 | 8.8 | 15.2 | 0.1 | 1.1 | 2.4 | 96.5 | 93.5 |
| | DT-L | 98.5 | 99.5 | 8.7 | 15.2 | 0.1 | 1.2 | 2.9 | 98.3 | 94.6 |
| P4 | Clinic | *89.6* | 99.5 | *15.3* | *27.6* | 0.1 | 0.8 | 2.2 | 90.3 | 87.4 |
| | DT-H | 93.8 | 99.2 | 13.5 | 23.4 | 0.0 | 0.3 | 1.2 | 91.4 | 87.4 |
| | DT-L | 17.9 | 95.7 | 14.2 | 24.8 | 0.0 | 0.8 | 2.7 | 94.7 | 92.0 |
| P5 | Clinic | 99.5 | 99.3 | *12.7* | 23.1 | *2.9* | *4.9* | *7.1* | 95.5 | 94.2 |
| | DT-H | 96.1 | 99.9 | 8.3 | 17.2 | 0.7 | 2.8 | 4.8 | 91.6 | 91.1 |
| | DT-L | 97.9 | 100.0 | 8.7 | 16.8 | 2.0 | 4.0 | 6.0 | 92.1 | 91.6 |

**Table 4.** Dose volume endpoint comparisons of DIL, CTV, and OARs for prostate SBRT patients, obtained by online evaluation (Eval.) for the clinical plans (clinic) and the plans by the proposed DT framework based on the daily corrected CBCT acquired from the second treatment fraction images (CB2). Underlined numbers indicate dosimetric values that surpass the dose constraints outlined in Table 1, and can trigger online adaptation workflow.

| CB2 Eval. | TP | DIL (%) | CTV (%) | Bladder ($cm^3$) | | Rectum ($cm^3$) | | | Urethra (%) | |
|---|---|---|---|---|---|---|---|---|---|---|
| | | V100 | V100 | V20.8Gy | V14.6Gy | V23Gy | V17.6Gy | V100 | V100 | D10 |
| P1 | Clinic | 98.7 | 99.7 | 7.4 | 17.6 | *1.3* | 3.4 | 6.2 | 91.2 | 88.6 |
| | DT-H | 96.2 | 98.0 | 7.2 | 15.4 | 1.2 | 3.8 | 7.3 | 90.0 | 86.7 |
| | DT-L | 97.3 | 98.4 | 7.6 | 16.2 | 1.0 | 3.6 | 7.7 | 90.5 | 88.1 |
| P2 | Clinic | 100.0 | 98.8 | *23.6* | *37.7* | 0.4 | 1.4 | 2.7 | 94.6 | 91.7 |
| | DT-H | 99.9 | 99.5 | 21.7 | 35.6 | 0.1 | 0.8 | 3.0 | 94.2 | 92.5 |
| | DT-L | 98.2 | 99.8 | 20.9 | 33.9 | 0.4 | 2.0 | 3.7 | 94.3 | 92.0 |
| P3 | Clinic | 100.0 | 99.9 | *13.6* | 24.1 | 1.0 | 3.5 | 5.9 | 105.0 | *102.4* |
| | DT-H | 99.8 | 100.0 | 9.5 | 16.8 | 0.8 | 2.3 | 3.8 | 100.6 | 96.7 |
| | DT-L | 99.7 | 100.0 | 9.4 | 17.0 | 0.8 | 2.1 | 3.7 | 100.2 | 96.6 |
| P4 | Clinic | 100.0 | 99.6 | *13.0* | *28.8* | 1.0 | 3.4 | 6.7 | 98.4 | *96.6* |
| | DT-H | 98.2 | 99.6 | 12.3 | 26.1 | 0.5 | 2.8 | 5.9 | 89.8 | 88.3 |
| | DT-L | 99.5 | 99.6 | 9.4 | 21.6 | 0.5 | 2.6 | 5.7 | 90.2 | 88.4 |
| P5 | Clinic | 98.6 | 99.9 | *14.7* | *26.9* | *1.9* | *4.3* | 6.6 | 91.8 | 91.2 |
| | DT-H | 99.7 | 100.0 | 9.8 | 19.5 | 0.9 | 2.5 | 4.5 | 91.0 | 90.2 |
| | DT-L | 99.1 | 100.0 | 9.8 | 19.7 | 1.7 | 3.8 | 5.8 | 91.0 | 89.9 |

### 3.3 Online adaptative treatment reoptimization using prior planning conditions

This section aims to demonstrate that DT-based plans can attain comparable quality to clinical plans while reducing reoptimization time. Table 5 outlines the online reoptimization (REopt) outcomes using daily CBCT images from Fx1, encompassing five plans per patient. The DT-H-REopt-A represents the baseline DT-based reoptimized plan, utilizing the pCT with the highest image similarity composite score to the daily CBCT, as defined in Table 2, with "A" indicating reoptimization within the 10-minute online adaptive planning threshold. The DT-L-REopt-A and DT-L-REopt-B denote DT-based reoptimized plans using the pCT with the lowest similarity score from Table 2, where "B" signifies a longer reoptimization duration yet comparable plan quality to the DT-H-REopt-A baseline. The Clinic-REopt-A and Clinic-REopt-B reflect the initial clinical plan, derived from the TPCT and reoptimized with daily CBCT. The composite plan quality scores, derived from dose statistics and all the scoring functions in Figure 2, increase with improved quality. Table 5 indicates that, under equivalent reoptimization time, DT-H-REopt-A yields the highest plan quality. With extended optimization, DT-L and clinical plans approach the quality score of DT-H plans. Table 6 corroborates these findings for DT-H-REopt-A plan qualities assessed with CB2 from Fx2, except for Patient 4, where DT-L-REopt-A achieves a marginally higher score (0.1 above DT-H-REopt-A).

Tables 5-6 demonstrate that DT-H-REopt-A achieves a mean reoptimization time of 5.5 ± 2.7 minutes with a plan quality score of 157.2 ± 5.6. In contrast, DT-L-REopt-B and Clinic-REopt-B require 14.6 ± 9.1 and 19.8 ± 11.9 minutes, respectively, with plan quality scores of 155.1 ± 6.0 and 153.8 ± 6.0, to achieve comparable plan quality to DT-H-REopt-A. For dose statistics reoptimized on daily CBCT, DT-H-REopt-A delivers V100 coverage of 99.5% ± 0.6% for the DIL and 99.8% ± 0.2% for the CTV. For OARs, DT-H-REopt-A achieves a bladder V20.8Gy of 11.4 ± 4.2 $cm^3$, rectum V23Gy of 0.7 ± 0.4 $cm^3$, and urethra D10 of 90.9% ± 2.3%. Figures 4 and 5 demonstrate that, given adequate time, all DT-based and clinical plans can reach comparable quality by dose-volume histogram (DVH) comparisons. However, the clinical and DT-L plans, generated from the least similar pCT image set, often exceeds the 10-minute replanning constraint for proton online adaptive treatment.

**Table 5.** Dose volume endpoint comparisons of dose statistics, obtained by online reoptimization (REopt) for the clinical plans (clinic) and the plans by the proposed DT framework based on the daily corrected CBCT acquired from the first treatment fraction images (CB1). Underlined numbers indicate the highest plan quality scores for each patient.

| **CB1 REopt** | | DIL (%) | CTV (%) | Bladder (cc) | | Rectum (cc) | | | Urethra (%) | | REopt Time (min) | Plan Quality Score |
|---|---|---|---|---|---|---|---|---|---|---|---|---|
| | | V100 | V100 | V20.8Gy | V14.6Gy | V23Gy | V17.6Gy | V13Gy | D0.03cc | D10 | | |
| P1 | DT-H-REopt-A | 98.5 | 100.0 | 10.0 | 19.8 | 1.1 | 2.8 | 5.3 | 92.0 | 89.3 | 9.0 | ***154.3*** |
| | DT-L-REopt-A | 98.3 | 99.5 | 10.4 | 21.9 | 1.1 | 3.6 | 6.7 | 92.2 | 90.6 | 9.1 | 150.8 |
| | DT-L-REopt-B | 98.1 | 99.7 | 10.0 | 21.5 | 1.1 | 3.5 | 6.5 | 94.1 | 91.6 | 19.0 | 150.4 |
| | Clinic-REopt-A | 98.1 | 99.8 | 9.7 | 22.2 | 1.2 | 3.3 | 5.9 | 93.4 | 92.2 | 9.7 | 150.0 |
| | Clinic-REopt-B | 98.1 | 99.8 | 9.3 | 21.3 | 1.2 | 3.2 | 5.8 | 91.5 | 89.2 | 28.0 | 151.5 |
| P2 | DT-H-REopt-A | 99.8 | 99.7 | 18.6 | 33.5 | 1.0 | 3.0 | 5.4 | 94.9 | 93.8 | 8.3 | ***147.2*** |
| | DT-L-REopt-A | 99.3 | 99.8 | 17.8 | 32.3 | 1.1 | 3.4 | 6.5 | 94.7 | 93.5 | 7.8 | 145.5 |
| | DT-L-REopt-B | 99.7 | 99.9 | 18.1 | 31.2 | 1.1 | 3.4 | 6.6 | 94.6 | 92.3 | 31.3 | 147.6 |
| | Clinic-REopt-A | 97.9 | 99.5 | 17.4 | 31.6 | 1.0 | 3.1 | 5.7 | 94.7 | 93.1 | 9.0 | 142.3 |
| | Clinic-REopt-B | 99.7 | 99.6 | 17.6 | 31.2 | 1.1 | 3.6 | 6.8 | 93.8 | 91.1 | 46.0 | 148.2 |
| P3 | DT-H-REopt-A | 100.0 | 99.9 | 9.1 | 15.5 | 0.1 | 1.1 | 2.4 | 95.8 | 94.0 | 4.7 | ***164.0*** |
| | DT-L-REopt-A | 100.0 | 100.0 | 9.0 | 15.4 | 0.2 | 1.2 | 3.0 | 97.9 | 94.7 | 4.6 | 162.9 |
| | DT-L-REopt-B | 100.0 | 100.0 | 9.0 | 15.5 | 0.2 | 1.2 | 3.0 | 97.8 | 94.5 | 16.4 | 163.1 |
| | Clinic-REopt-A | 100.0 | 99.9 | 11.9 | 21.5 | 0.2 | 2.1 | 4.4 | 98.2 | 95.9 | 4.5 | 158.6 |
| | Clinic-REopt-B | 100.0 | 100.0 | 11.8 | 21.5 | 0.2 | 2.1 | 4.3 | 98.6 | 96.9 | 18.1 | 157.8 |
| P4 | DT-H-REopt-A | 99.8 | 99.2 | 9.2 | 19.4 | 0.0 | 0.4 | 1.4 | 91.1 | 88.3 | 2.4 | ***165.9*** |
| | DT-L-REopt-A | 84.7 | 94.9 | 11.0 | 21.8 | 0.1 | 1.0 | 2.9 | 96.9 | 92.9 | 2.4 | 113.4 |
| | DT-L-REopt-B | 99.8 | 99.3 | 6.3 | 15.0 | 0.2 | 1.9 | 4.6 | 90.4 | 88.0 | 6.4 | 164.5 |
| | Clinic-REopt-A | 99.2 | 98.7 | 10.1 | 22.7 | 0.1 | 0.8 | 2.3 | 93.2 | 90.5 | 2.9 | 161.1 |
| | Clinic-REopt-B | 99.9 | 99.3 | 5.2 | 14.8 | 0.3 | 1.6 | 3.8 | 88.8 | 86.2 | 9.8 | 166.5 |
| P5 | DT-H-REopt-A | 99.0 | 100.0 | 9.2 | 18.0 | 1.0 | 3.2 | 5.3 | 91.1 | 90.8 | 3.8 | ***156.2*** |
| | DT-L-REopt-A | 99.1 | 100.0 | 8.9 | 17.0 | 2.0 | 4.0 | 6.0 | 92.6 | 92.0 | 3.6 | 149.3 |
| | DT-L-REopt-B | 99.5 | 99.9 | 9.3 | 17.7 | 1.9 | 3.8 | 5.7 | 91.2 | 90.6 | 6.7 | 152.4 |
| | Clinic-REopt-A | 99.5 | 99.9 | 12.0 | 22.3 | 2.1 | 4.5 | 6.8 | 91.8 | 91.0 | 2.9 | 141.5 |
| | Clinic-REopt-B | 99.5 | 100.0 | 12.2 | 22.4 | 2.0 | 4.2 | 6.2 | 90.2 | 89.3 | 6.1 | 147.7 |

**Table 6.** Dose volume endpoint comparisons of dose statistics, obtained by online reoptimization (REopt) for the clinical plans (clinic) and the plans by the proposed DT framework based on the daily corrected CBCT acquired from the second treatment fraction images (CB2). Underlined numbers indicate the highest plan quality scores for each patient.

| **CB2 REopt** | | DIL (%) | CTV (%) | Bladder (cc) | | Rectum (cc) | | | Urethra (%) | | REopt Time (min) | Plan Quality Score |
|---|---|---|---|---|---|---|---|---|---|---|---|---|
| | | V100 | V100 | V20.8Gy | V14.6Gy | V23Gy | V17.6Gy | V13Gy | D0.03cc | D10 | | |
| | DT-H-REopt-A | 98.5 | 99.9 | 8.4 | 16.3 | 1.0 | 3.4 | 6.6 | 92.1 | 90.4 | 3.2 | ***154.1*** |
| | DT-L-REopt-A | 97.0 | 99.8 | 8.2 | 16.6 | 1.5 | 4.1 | 7.8 | 94.6 | 91.7 | 3.0 | 133.4 |
| P1 | DT-L-REopt-B | 97.8 | 99.7 | 8.6 | 17.0 | 1.4 | 3.5 | 6.5 | 95.4 | 92.0 | 8.5 | 148.6 |
| | Clinic-REopt-A | 98.8 | 99.9 | 8.2 | 18.6 | 1.4 | 3.5 | 6.4 | 91.8 | 88.9 | 3.9 | 153.1 |
| | Clinic-REopt-B | 97.7 | 100.0 | 8.7 | 19.0 | 1.4 | 3.4 | 6.3 | 91.3 | 88.2 | 22.2 | 149.4 |
| | DT-H-REopt-A | 100.0 | 100.0 | 19.9 | 33.6 | 0.4 | 1.4 | 4.1 | 92.9 | 91.6 | 6.6 | ***151.8*** |
| | DT-L-REopt-A | 100.0 | 100.0 | 18.9 | 32.6 | 0.5 | 2.3 | 4.6 | 93.9 | 91.9 | 6.8 | 151.4 |
| P2 | DT-L-REopt-B | 100.0 | 99.9 | 18.3 | 31.8 | 0.5 | 2.4 | 4.9 | 93.8 | 92.0 | 14.8 | 151.8 |
| | Clinic-REopt-A | 100.0 | 99.8 | 20.6 | 34.1 | 0.4 | 1.9 | 3.9 | 93.8 | 91.0 | 6.7 | 150.7 |
| | Clinic-REopt-B | 100.0 | 99.8 | 19.6 | 33.5 | 0.5 | 2.1 | 4.3 | 93.8 | 91.1 | 26.5 | 151.2 |
| | DT-H-REopt-A | 99.9 | 99.8 | 9.5 | 16.8 | 0.9 | 2.4 | 4.0 | 100.6 | 93.5 | 9.7 | ***158.6*** |
| | DT-L-REopt-A | 99.7 | 99.9 | 9.6 | 17.2 | 0.9 | 2.3 | 3.7 | 93.6 | 98.9 | 10.5 | 155.4 |
| P3 | DT-L-REopt-B | 99.9 | 99.7 | 9.6 | 17.0 | 0.9 | 2.3 | 3.8 | 100.0 | 94.3 | 28.1 | 158.5 |
| | Clinic-REopt-A | 100.0 | 99.9 | 13.2 | 23.7 | 1.0 | 3.6 | 5.8 | 100.6 | 95.9 | 9.5 | 153.0 |
| | Clinic-REopt-B | 99.9 | 99.7 | 12.8 | 23.2 | 1.1 | 3.6 | 6.0 | 101.5 | 94.8 | 19.9 | 153.0 |
| | DT-H-REopt-A | 99.8 | 99.6 | 9.8 | 23.3 | 0.6 | 2.9 | 6.0 | 90.3 | 88.1 | 3.3 | 160.0 |
| | DT-L-REopt-A | 99.3 | 99.9 | 8.6 | 20.6 | 0.5 | 2.7 | 5.8 | 88.9 | 86.6 | 3.2 | ***160.1*** |
| P4 | DT-L-REopt-B | 99.0 | 99.4 | 8.6 | 20.2 | 0.6 | 2.9 | 5.9 | 87.4 | 85.1 | 7.7 | 159.4 |
| | Clinic-REopt-A | 100.0 | 100.0 | 9.2 | 24.1 | 0.7 | 3.2 | 6.3 | 95.1 | 92.6 | 3.6 | 158.3 |
| | Clinic-REopt-B | 99.9 | 100.0 | 9.3 | 22.7 | 0.8 | 3.1 | 6.2 | 89.1 | 87.5 | 9.1 | 160.5 |
| | DT-H-REopt-A | 99.6 | 100.0 | 9.9 | 19.7 | 1.0 | 2.6 | 4.6 | 90.6 | 89.1 | 4.0 | ***159.4*** |
| | DT-L-REopt-A | 99.6 | 100.0 | 10.0 | 19.8 | 1.6 | 3.7 | 5.8 | 91.0 | 90.2 | 4.0 | 154.2 |
| P5 | DT-L-REopt-B | 99.4 | 99.9 | 10.2 | 20.3 | 1.4 | 3.4 | 5.7 | 91.0 | 90.6 | 7.3 | 154.4 |
| | Clinic-REopt-A | 99.0 | 99.9 | 14.2 | 26.3 | 1.8 | 4.2 | 6.6 | 91.1 | 90.5 | 5.2 | 144.0 |
| | Clinic-REopt-B | 99.5 | 100.0 | 12.9 | 24.8 | 1.4 | 3.7 | 6.1 | 89.9 | 88.9 | 12.2 | 152.6 |

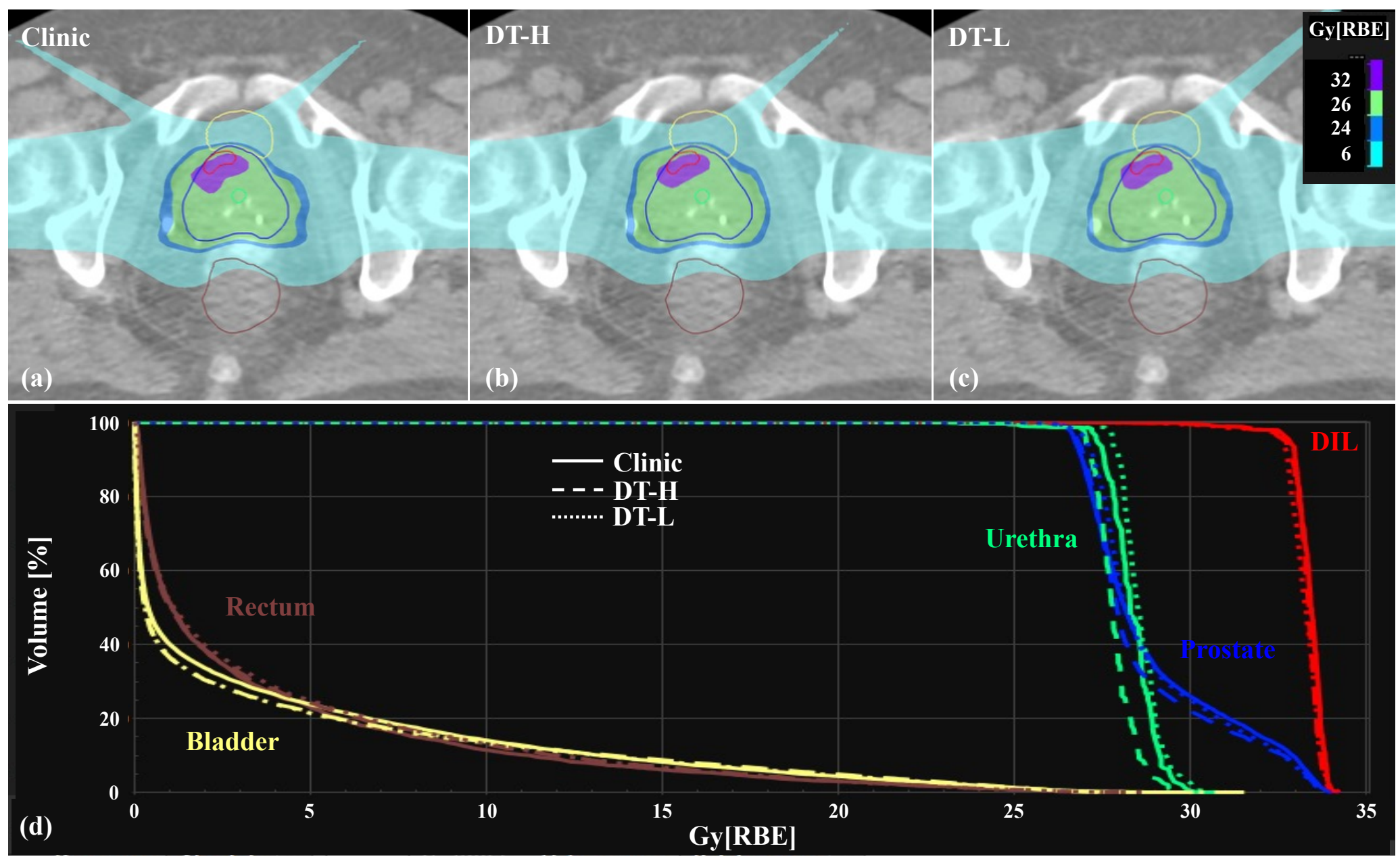

**Figure 4**. Dosimetry comparisons of dose evaluation on the daily CBCT for Patient 1 using the clinical, DT-H and DT-L plans. The DT-H and DT-L plans were created using the pCT generated from the DT framework, where H and L denotes high and low similarities between the pCT and the daily CBCT. The dose distributions in color wash for the (a) clinical, (b) DT-H, and (c) DT-L plan, with contours of DIL (red), CTV (blue), bladder (yellow), urethra (green), and rectum (deep reddish-brown). (d) The dose-volume histogram (DVH) for DIL, CTV, and OAR structures, where the solid, dashed, and dotted lines represent the clinical, DT-H, and DT-L plans.

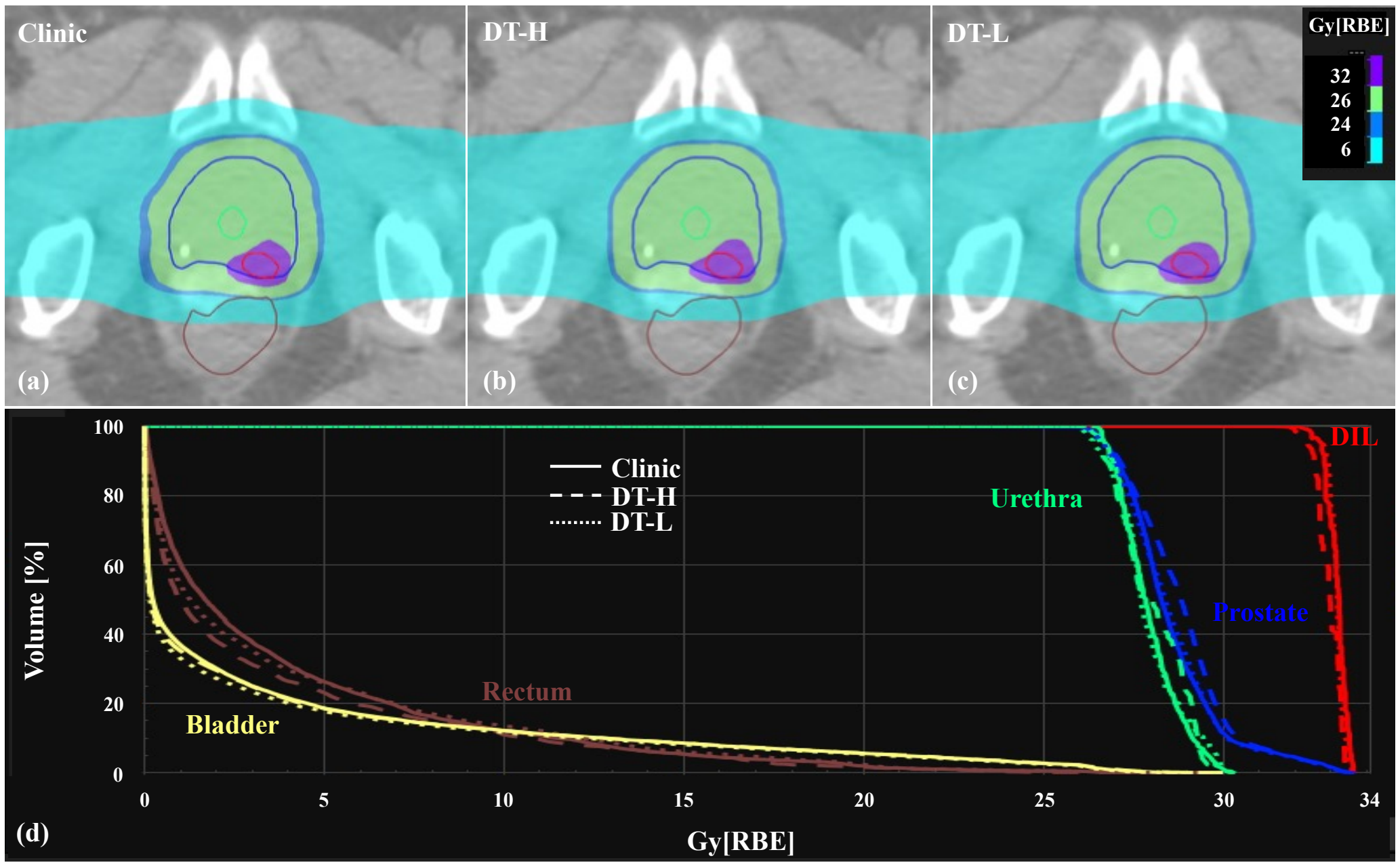

**Figure 5**. Dosimetry comparisons of dose evaluation on the daily CBCT for Patient 2 using the clinical, DT-H and DT-L plans. The DT-H and DT-L plans were created using the pCT generated from the DT framework, where H and L denotes high and low similarities between the pCT and the daily CBCT. The dose distributions in color wash for the (a) clinical, (b) DT-H, and (c) DT-L plan, with contours of DIL (red), CTV (blue), bladder (yellow), urethra (green), and rectum (deep reddish-brown). (d) The dose-volume histogram (DVH) for DIL, CTV, and OAR structures, where the solid, dashed, and dotted lines represent the clinical, DT-H, and DT-L plans.

## 4 Discussion

We demonstrated a DT framework with integration of a DL-based DIR methodology using an extensive prior radiation treatment database to enable rapid online adaptive treatment optimization. This data-driven approach ensures that DIL coverage remains equivalent to that of traditional clinical plans, while significantly reducing the reliance on time-consuming replanning procedures due to insufficient target coverages or OAR toxicity. In the existing clinical workflow, replanning becomes imperative when anatomical variations lead to insufficient DIL coverage or excessive doses to critical organs, a process that demands a five-day turnaround for SBRT protocols. During this period, patients are typically treated with the original plan until the updated plan is available. However, in specific clinical scenarios, such as SBRT treatments, conventional therapies with only a few fractions remaining, or cases where a physician explicitly requests a pause, treatment is halted until a revised plan can be implemented. In stark contrast, the DT framework proactively generates a diverse array of treatment plans using pCT images, harnessing population-based insights from prior treatment data to anticipate potential anatomical changes. These pre-generated plans serve as robust preconditions, facilitating swift online reoptimization tailored to individual patient needs. The efficacy of this approach is substantiated by Tables 5 and 6, which highlight that DT-based plans consistently achieve superior quality scores and can be completed within 5.5 ± 2.7 minutes achieving a plan quality score of 157.2 ± 5.6, thereby establishing a foundation for the practical implementation of real-time adaptive treatment strategies. By effectively addressing interfractional anatomy changes and fast online plan reoptimization, this methodology not only holds significant potential for enhancing treatment outcomes but also for enhancing healthy tissue sparing, ultimately achieving precision prostate cancer treatment for each individual patient.

Initial CT-based treatment plans typically achieve effective target coverage and OAR sparing. However, a 2- to 3-week clinical workflow between the initial CT scan and the first treatment day can lead to anatomical discrepancies. These may arise from variations in bladder filling, weight loss, or setup uncertainties, potentially affecting the accuracy of the initial plan. Table 3 underscores a critical challenge posed by intrafractional anatomical variability on treatment day, as evidenced by Patient 2 (P2) and Patient 4 (P4). For P2, CBCT-based dose evaluation of the clinical plan showed bladder V20.8Gy and V14.6Gy values of 26.5 $cm^3$ and 41.7 $cm^3$, respectively, surpassing the constraints of 10 $cm^3$ and 25 $cm^3$ specified in Table 1. This case illustrates a scenario where initial planning, Figure 6(a1)-(a3), demonstrated clear separation between the bladder and prostate, with a bladder volume of 226 cc. However, daily CBCT imaging, Figure 6(b1)-(b3), on treatment day revealed a significant bladder volume increase to 340 cc, positioning it closer to high-dose treatment regions and causing substantial bladder overdose. Additionally, these anatomical changes led to significant underdosing of the clinical plan's DIL and CTV, with V100 values reduced to 89.7% and 93.3%, respectively. This scenario highlights the necessity of adaptive strategies, where online reoptimization, Figure 6(c1)-(c3), proves essential in addressing such challenges. The DT-H plans, leveraging accurate predictions of patient anatomical motion, achieve significantly improved reoptimized plan quality in under 10 minutes. This efficiency stems from the high-fidelity prior planning conditions inherent to DT-H plans, as substantiated by the comprehensive data in Table 5. In contrast, P4 in Table 3 presents a scenario where the DIL is well-covered during initial planning, as shown in Figure 7(a1)-(a3), but the DT-L plan fails to achieve adequate DIL coverage during plan evaluation with daily CBCT, as depicted in Figure 7(b3). This shortfall is attributed to the low similarity between the planned and actual daily anatomical conditions reflected in the CBCT images. Figure 7(c3) demonstrates that rapid online reoptimization effectively restores DIL coverage. Collectively, these findings confirm that DT-based plans, characterized by high anatomical similarity, enable fast online reoptimization (<10 minutes), as illustrated in Tables 5-6, while maintaining clinically comparable quality. This positions DT-based plans as a viable advancement in adaptive radiotherapy.

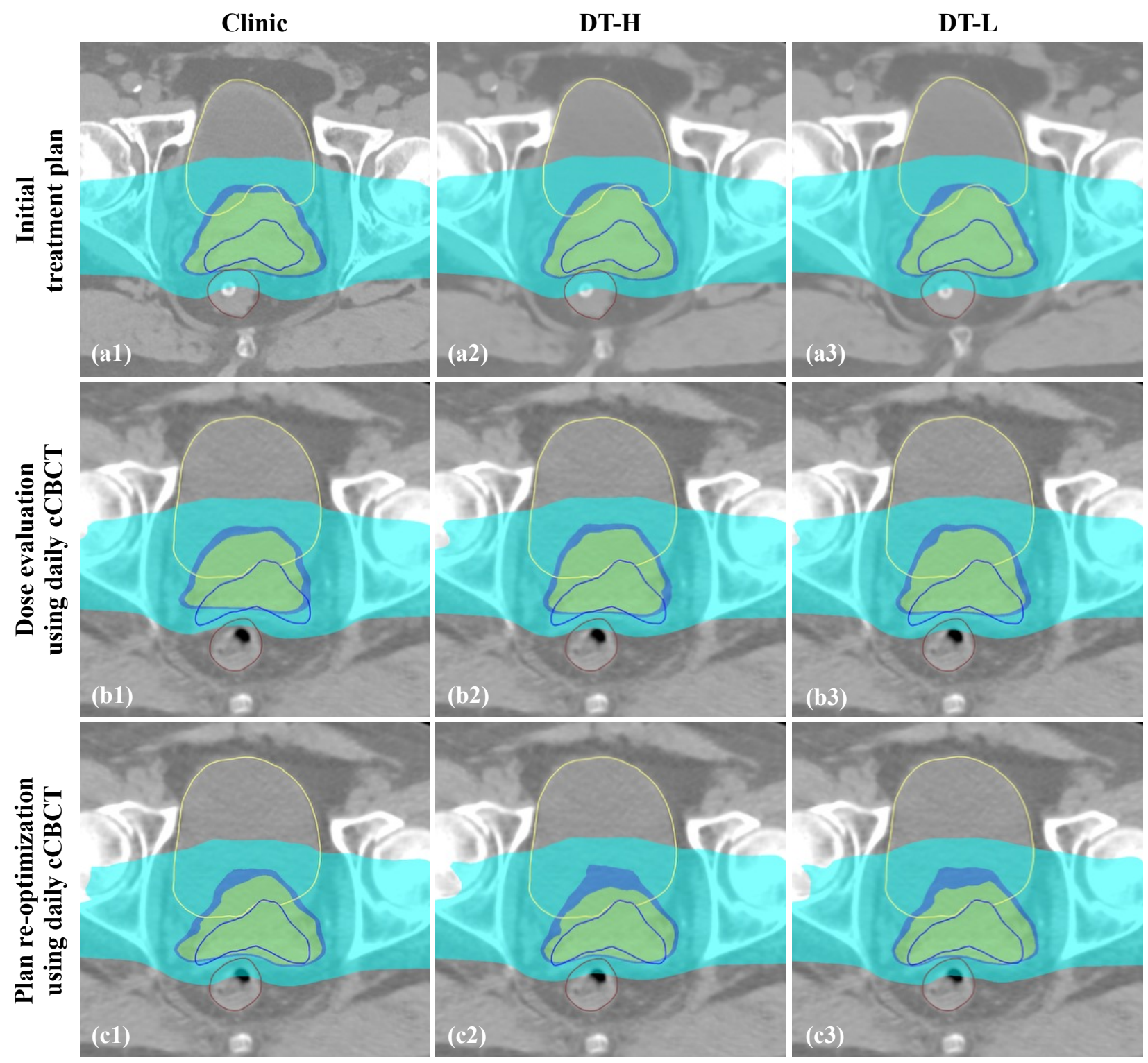

**Figure 6**. Dosimetry comparisons of (a1-a3) initial planning doses, (b1-b3) daily CBCT evaluation doses, (c1-c3) plan reoptimized doses for (1) clinic, (2) DT-H, and (3) DT-L plans for Patient 2 (P2).

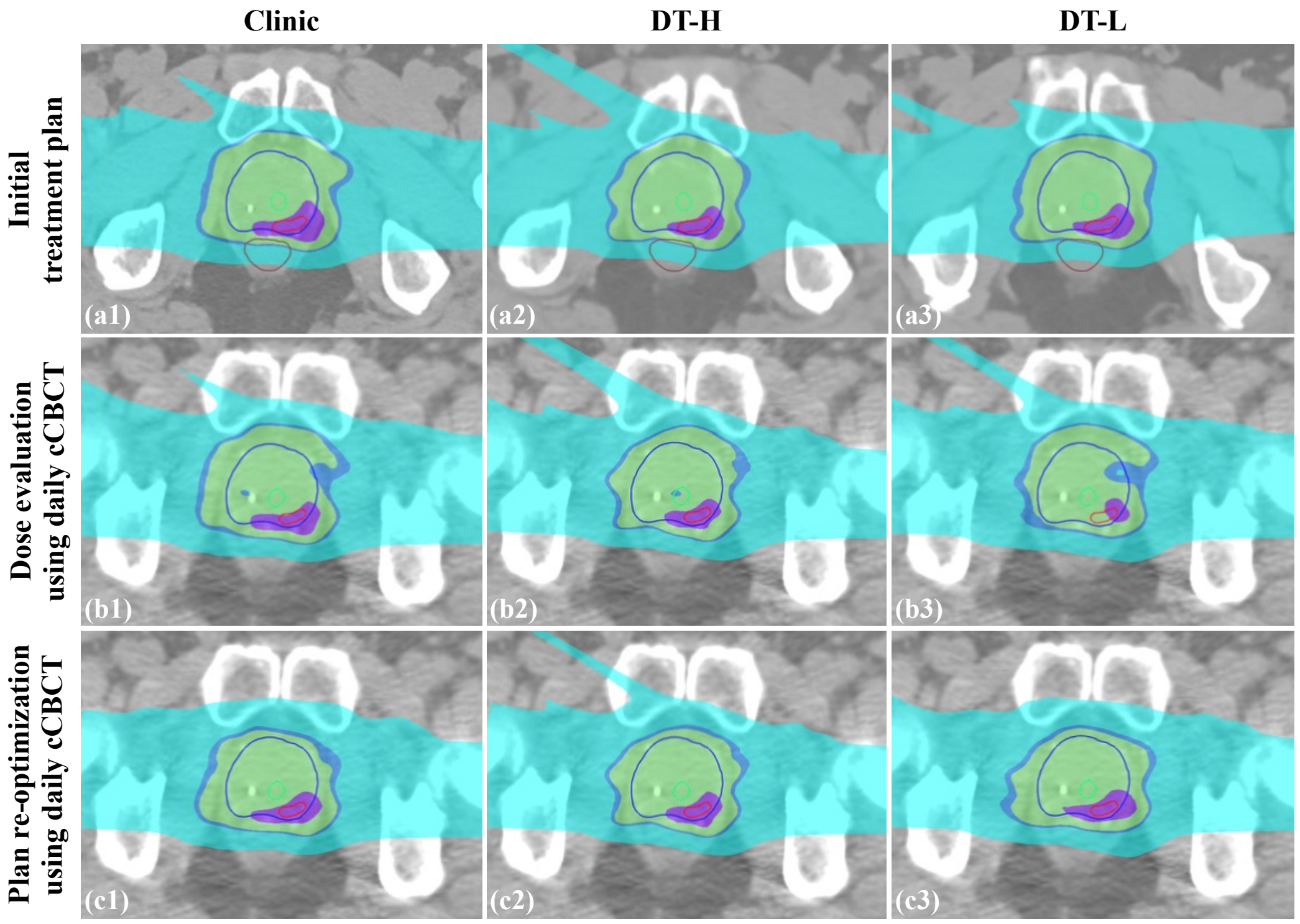

**Figure 7**. Dosimetry comparisons of (a1-a3) initial planning doses, (b1-b3) daily CBCT evaluation doses, (c1-c3) plan reoptimized doses for (1) clinic, (2) DT-H, and (3) DT-L plans for Patient 4 (P4).

The overarching objective of the DT framework is to achieve fast personalized radiation therapy, raising two critical questions to enhance its practical feasibility: (1) How can online plan reoptimization be efficiently and reliably performed within 10 minutes using automation to improve accuracy and consistency? (2) How can the inherent uncertainty in patient anatomy be systematically and effectively quantified to inform treatment adjustments? Reinforcement learning (RL) (Zhang *et al.*, 2021) and deep reinforcement learning (DRL) show promise for photon-based prostate cancer treatment planning, as demonstrated by recent studies (Hrinivich and Lee, 2020; Shen *et al.*, 2020; Shen *et al.*, 2021). However, implementing DRL within a commercial treatment planning system can require up to 2 hours for plan optimization (Gao *et al.*, 2023). With effective initial plan conditions, Hrinivich *et al.* (Hrinivich *et al.*, 2024) showed that DRL can rapidly generate plans for photon-based prostate cancer treatment planning. The current DT framework prioritizes the delivery of prescribed radiation doses to the DIL while simultaneously striving to minimize exposure to surrounding healthy tissues. An additional layer of complexity arises from proton range uncertainty, estimated at 3.5%, which stems from limitations in CT-based material characterization techniques and may compromise the conformal precision of proton therapy. To address this, future research endeavors are poised to investigate advanced DL-based methodologies for material characterizations (Chang *et al.*, 2022a; Chang *et al.*, 2022b) and advanced volumetric generation without motion artifacts (Shen *et al.*, 2019; Pan *et al.*, 2024; Pan *et al.*, 2025), aiming to reduce this uncertainty and further limit irradiation to normal tissues. This integration will accommodate the growing complexity of the system and enhance data accessibility, ensuring that the DT framework can adapt dynamically to diverse patient profiles and clinical scenarios, thereby maximizing its therapeutic potential.

## 5 Conclusions

We developed a novel framework leveraging DT concepts to achieve fast adaptive proton prostate SBRT with DIL boost. This approach optimizes DT-derived treatment plans by incorporating DL-based DIR to enable treatment planning using potential patients' interfractional anatomy changes based on prior treatment knowledge. Plan assessments, conducted using CBCT, revealed that DT-based plans provide comparable or improved DIL coverages with clinical equivalent quality while substantially sparing bladder, rectum, and urethra, thereby potentially mitigating toxicity risks. By integrating the DT paradigm, this adaptive proton therapy framework offers significant potential to support clinical decision-making by optimizing treatment protocols for cancer radiotherapy.

## Acknowledgments

This research is supported in part by the National Institutes of Health under Award Number R01CA215718, and R01CA272991.

## Conflict of interest

The authors have no conflict of interests to disclose.

**Ethical Statement**

Emory IRB review board approval was obtained (IRB #114349), and informed consent was not required for this Health Insurance Portability and Accountability Act (HIPAA) compliant retrospective analysis. The research was conducted in accordance with the principles embodied in the Declaration of Helsinki and in accordance with local statutory requirements.

**Data availability statement**

The data cannot be made publicly available upon publication because they contain sensitive personal information. The data that support the findings of this study are available upon reasonable request from the authors.